\documentclass[fleqn,10pt]{wlscirep}
\usepackage[utf8]{inputenc}
\usepackage[T1]{fontenc}
\usepackage{bm}
\usepackage{lineno}

\title{Inferring the ecology of north-Pacific albacore tuna from catch-and-effort data}

\author[1,*]{Hirotaka Ijima}
\author[2]{Carolina Minte-Vera}
\author[3]{Yi-Jay Chang}
\author[1]{Daisuke Ochi}
\author[1]{Yuichi Tsuda}
\author[1]{Marko Jusup}

\affil[1]{Fisheries Resources Research Institute, Fisheries Research Agency, Yokohama 236-8648, Japan}
\affil[2]{Inter-American Tropical Tuna Commission, La Jolla, CA 92037, USA}
\affil[3]{Institute of Oceanography, National Taiwan University, Taipei 106, Taiwan}
\affil[*]{ijima\_hirotaka69@fra.go.jp}
\begin{abstract}
Catch-and-effort data are among the primary sources of information for assessing the status of terrestrial wildlife and fish. In fishery science, elaborate stock-assessment models are fitted to such data in order to estimate fish-population sizes and guide management decisions. Given the importance of catch-and-effort data, we scoured a comprehensive dataset pertaining to albacore tuna (\textit{Thunnus alalunga}) in the north Pacific ocean for novel ecological information content about this commercially valuable species. Specifically, we used unsupervised learning based on finite mixture modelling to reveal that the north Pacific albacore-tuna stock can be divided into four pseudo-cohorts ranging in age from approximately 3 to 12 years old. We discovered that smaller size pseudo-cohorts inhabit relatively high---subtropical to temperate---latitudes, with hotspots off the coast of Japan. Larger size pseudo-cohorts inhabit lower---tropical to subtropical---latitudes, with hotspots in the western and central north Pacific. These results offer evidence that albacore tuna prefer different habitats depending on their size and age, and point to long-term migratory routes for the species that the current tagging technology is unlikely to capture in full. We discuss the implications of the results for data-driven modelling of albacore tuna in the north Pacific, as well as the management of the north Pacific albacore-tuna fishery.
\end{abstract}
\begin{document}

\flushbottom
\maketitle

\thispagestyle{empty}

%\linenumbers

\section*{Introduction}

The importance of biological resources to humankind is hard to overstate, especially so in the context of food security. And yet, we as a species have time and again proven to be poor stewards of nature's riches, a state of affairs that has succinctly been captured in the phrase `the tragedy of the commons'~\cite{hardin1998extensions, hardin1968tragedy}. To exemplify, data compiled by the Food and Agriculture Organisation of the United Nations show that about 25\,\% of global fisheries collapsed in the period from 1950 to 2000~\cite{mullon2005dynamics}. Although prescriptions for avoiding the tragedy of the commons and ensuring sustainability have been documented~\cite{ostrom2009general, pretty2003social}, they are not panaceas~\cite{battersby2017can}. Behavioural patterns of overexploitation have proven remarkably robust across modern cultures~\cite{jusup2020behavioural}, adding to the evidence that achieving sustainability is an uphill battle.

In search of sustainability, science has devised a plethora of methods ranging in sophistication from simple rules of thumb to data-intensive quantitative models with mechanistic underpinning~\cite{weinbaum2013searching}. Collecting catch-and-effort data has for the longest time been one of the go-to methods for scientists, increasingly as an input into said quantitative models, but oftentimes also as a source of standalone indicators~\cite{weinbaum2013searching}. Catch per unit effort (CPUE) is one such indicator that has become a staple of wildlife~\cite{rist2010hunter, rist2008evaluating}, fishery~\cite{kuparinen2012increasing, maunder2004standardizing}, and even forestry~\cite{yousefpour2012review} related sustainability science. CPUE is often seen as a proxy for abundance although the relation between CPUE and abundance may in some instances be fairly complex~\cite{weinbaum2013searching}.

In modern fishery science, CPUE is the primary abundance indicator fed into sophisticated stock-assessment models such as Stock Synthesis 3~\cite{methot2013stock}. The models use CPUE on a per-fleet basis, with fleets being defined depending on their size selectivity as well as the country of origin. Size selectivity should in principle reflect the fishing gear employed, but this is often an oversimplifying assumption due to spatio-temporal patterns in the structure of fish stocks arising from, for example, migratory movement or variable recruitment~\cite{waterhouse2014using}. Attempts are therefore being made to leverage the spatio-temporal information contained in the CPUE itself~\cite{ichinokawa2010using}, or in the original catch-and-effort data, in order to organically arrive at data-driven fleet definitions. Motivated by an analogous line of reasoning, but shifting the focus towards deepening general ecological understanding, we explored the catch-and-effort data from the Japanese albacore-tuna fishery in the north Pacific ocean with the aim to organically arrive at a data-driven stock structure of these commercially exploited fish.

Albacore tuna (\textit{Thunnus alalunga}) is a migratory pelagic species in the family Scombridae that inhabits most tropical and temperate oceans across the globe. Adult individuals in the Pacific approximately grow up to 125\,cm fork length, 35\,kg body mass, and 21 years of age~\cite{wells2013age}. They mature at approximately 90\,cm fork length~\cite{juan2016global} and five years of age. This combination of life-history traits paired with the considerable commercial catch has led to some mentions of the species being vulnerable to overexploitation~\cite{nikolic2017review}, but albacore tuna is currently in the 'least concern' category of The International Union for Conservation of Nature (IUCN) Red list of Threatened Species. The list notes a declining population trend that is likely due to the size of albacore-tuna fishery. In 2018, this fishery yielded 235,000\,mt of landings, with a dock value of \$650\,m and a final value of \$2.1\,bn~\cite{mckinney2020netting}.

\begin{figure}[!t]
\centering
\includegraphics[scale=1.0]{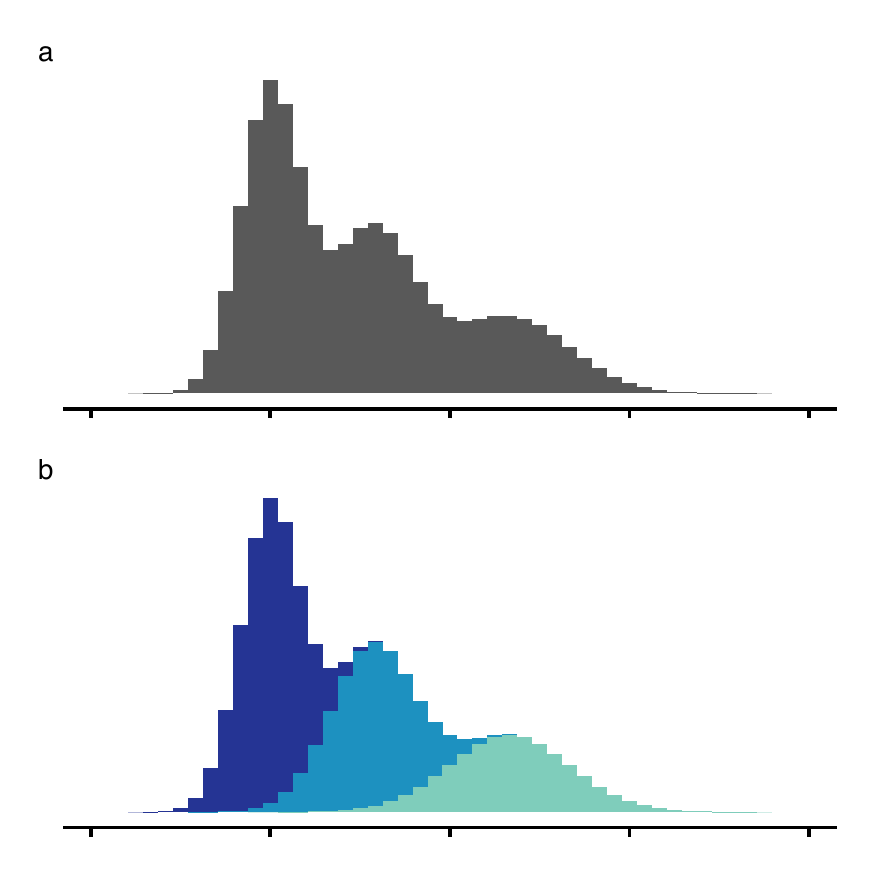}
\caption{\textbf{A schematic conveying the main idea behind unsupervised learning by means of finite mixture modelling.} \textbf{a}, Empirical probability density function derived from a dataset as a whole reveals a multimodal, mixture distribution. \textbf{b,} In many cases, multiple modes appear in composite datasets that can be broken down into more basic constituents. This is precisely the task of unsupervised learning with finite mixture modelling. Here, the original multimodal, mixture distribution is separated into three constituent, monomodal distributions.}
\label{f:sch}
\end{figure}

Given the value of the albacore-tuna fishery, as well as uncertainties surrounding the albacore-tuna stock structure~\cite{moore2020defining1, moore2020defining2}, we attempted extracting new ecological knowledge about the species from a previously unexplored perspective. Specifically, we applied unsupervised learning based on finite mixture modelling to a comprehensive catch-and-effort dataset recorded by the Japanese albacore-tuna longline fishery in the north Pacific ocean. This approach has the ability to separate a multimodal, mixture probability distribution into constituent, monomodal distributions~\cite{mclachlan2019finite, leisch2004flexmix}, which is potentially useful in the context of catch-and-effort data if catches originate from multiple, mutually distinct cohorts (Fig.~\ref{f:sch}). The dataset and the analyses are detailed in the Methods section. Briefly, the available dataset contained the number of fish caught, their mass, and the corresponding effort. Dividing the single-operation mass by the number of fish caught during the operation yielded the average mass, which served as input in the finite mixture model. The number of fish caught divided by the effort yielded CPUE, which served as an abundance indicator. We discovered that the north Pacific albacore tuna can reasonably be divided into four pseudo-cohorts, with the prefix `pseudo' being used to indicate `extracted' or `learnt' from data. Pseudo-cohorts exhibit different habitat preferences depending on size and age. Following the progression of habitat preferences throughout ontogeny, we could piece together a novel picture of migration pathways in the western north Pacific. We hereafter proceed to first describe and then discuss these results in some detail.

\section*{Results}

\begin{figure}[!t]
\centering
\includegraphics[scale=1.0]{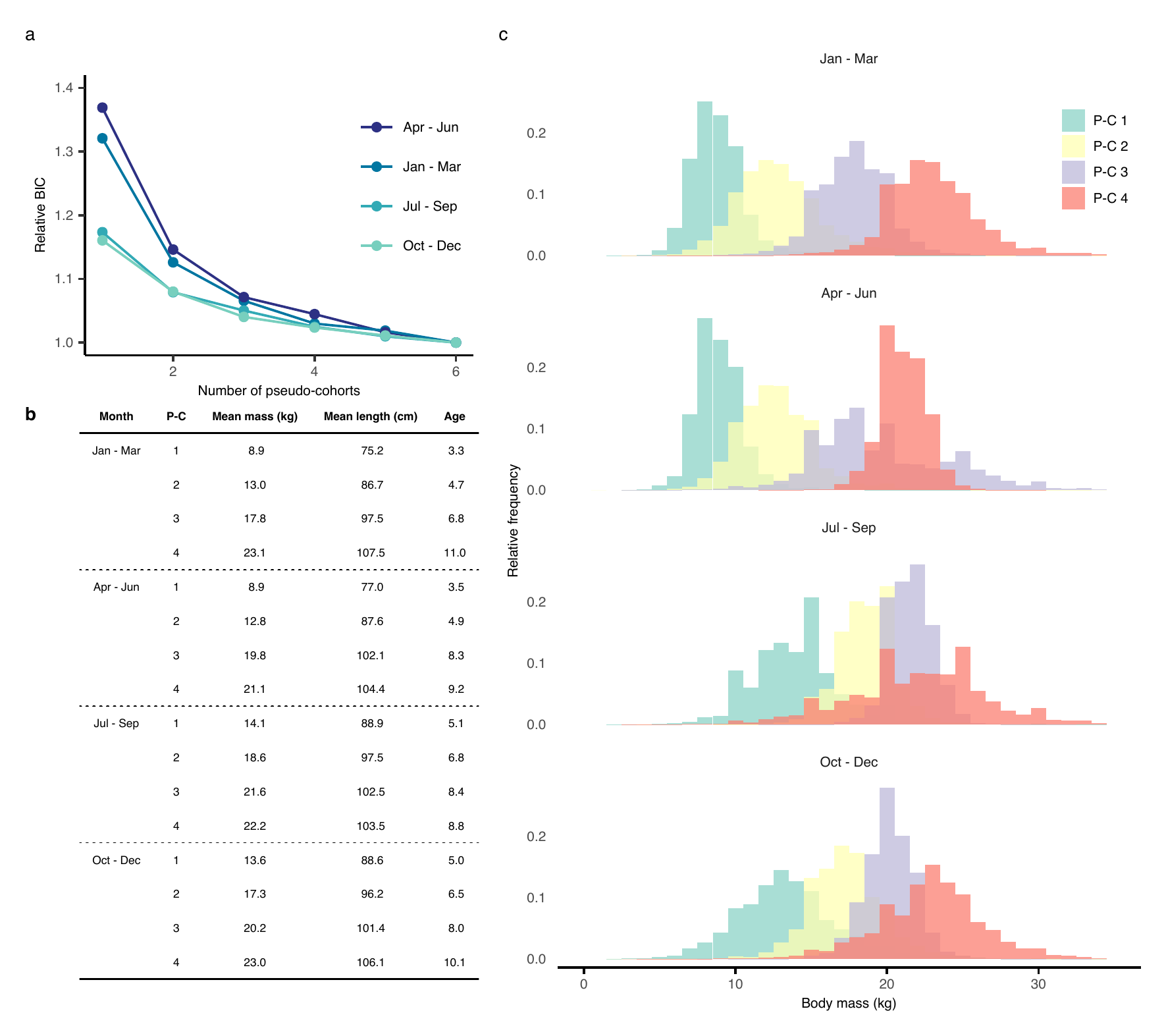}
\caption{\textbf{Division of seasonal catch data into four pseudo-cohorts by means of unsupervised learning based on finite mixture modelling.} \textbf{a,} A total of four pseudo-cohorts was chosen because the relative Bayesian information criterion (BIC) reveals limited marginal improvements by adding more pseudo-cohorts. \textbf{b,} Each pseudo-cohort is characterised by its mean body mass, as well as the corresponding estimates of mean body length and mean age. \textbf{c,} Breakdown of the multimodal, mixture probability distribution of the original dataset into four constituent, monomodal distributions, each of which defines a pseudo-cohort.}
\label{f:fmm}
\end{figure}

The initial step in finite mixture modelling is determining the likely number of constituent, monomodal distributions present in the multimodal, mixture probability distribution of the whole dataset. This number is treated as a free parameter whose value is chosen by the modeller, but with the help of some goodness-of-fit or information-criterion measure. Employing the relative Bayesian information criterion (BIC), we found that its marginal improvement is limited if the available catch-and-effort data are divided into more than four pseudo-cohorts (Fig.~\ref{f:fmm}a). We therefore decided to work with four pseudo-cohorts to avoid overfitting. Keeping the number of pseudo-cohorts constant throughout the year, despite having repeated the analysis for each quarter independently, further facilitates interpretability.

\begin{figure}[!t]
\centering
\includegraphics[scale=1.0]{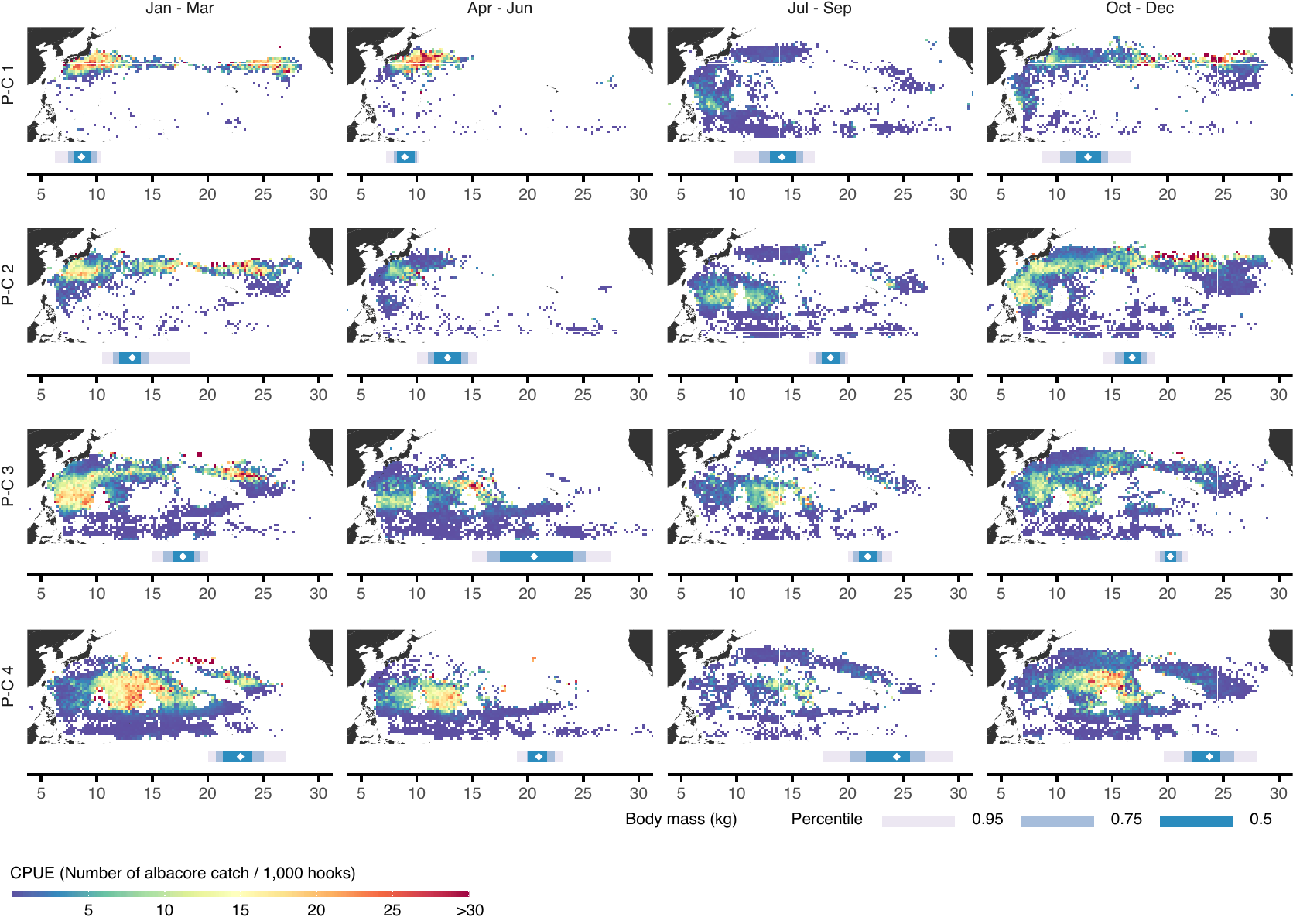}
\caption{\textbf{Catch per unit effort (CPUE) unveils the geographic and seasonal origins of each pseudo-cohort.} Shown underneath each pseudo-cohort's CPUE distribution are the corresponding mean body mass (diamond) and the 50th, 75th, and 95th body-mass percentile. The two smaller size pseudo-cohorts (1 \& 2) can mainly be found at relatively high---subtropical to temperate---latitudes, with hotspots off the coast of Japan in boreal winter and spring months. The two larger size pseudo-cohorts (3 \& 4) are more likely to be found at lower---tropical and subtropical---latitudes, with hotspots in the western north Pacific in boreal winter months and central north Pacific throughout much of the year.}
\label{f:cpue}
\end{figure}

We characterised pseudo-cohorts by their mean body mass (Fig.~\ref{f:fmm}b), which was estimated from learnt constituent, monomodal distributions (Fig.~\ref{f:fmm}c). We also computed for reference the corresponding mean body length and mean age from known weight-length and length-age relationships~\cite{nikolic2017review}. Pseudo-cohorts exhibit a couple of interesting properties that are best understood in conjunction with CPUE, whose role is to unveil pseudo-cohort geographic and seasonal origins (Fig.~\ref{f:cpue}). First, the distribution of body mass for the two smaller size pseudo-cohorts shifts substantially towards larger masses throughout the year (cf.\ upper vs.\ lower panels in Fig.~\ref{f:fmm}c). These pseudo-cohorts gather in high---subtropical to temperate---latitudes from January to June, with hotspots off the coast of Japan (Fig.~\ref{f:cpue}). Mean body mass approximately equals 9\,kg and 13\,kg, meaning that the fish are in an ontogenetic stage when growth is fast, which to some degree accounts for the upward shift in body-mass distribution. Additionally, the fish disperse from putative hotspots between July and September, and in part start appearing more southward where they intermix with fish from the two larger size pseudo-cohorts. For these reasons, mean body mass of the smaller size pseudo-cohorts appears larger in the second half of the year, approximately equalling 14\,kg and 18\,kg.

The second interesting property is that constituent, monomodal distributions for the two larger size pseudo-cohorts are sharply separated in the first quarter, but less so for the rest of the year, especially in the period between April and September (cf.\ top vs.\ other panels in Fig.~\ref{f:fmm}c). These pseudo-cohorts gather in lower---tropical to subtropical---latitudes, with hotspots in the western north Pacific in boreal winter months and central north Pacific throughout much of the year (Fig.~\ref{f:cpue}). The former hotspot attracts fish with mean body mass between approximately 18\,kg and 20\,kg, whereas the latter hotspot attracts the largest fish with mean body mass between 21\,kg and 23\,kg. In particular, the fish disperse from the hotspot in the western north Pacific, and start appearing more centrally where intermixing occurs. The distinction between the two larger size pseudo-cohorts somewhat blurs due to this intermixing.

\begin{figure}[!t]
\centering
\includegraphics[scale=1.0]{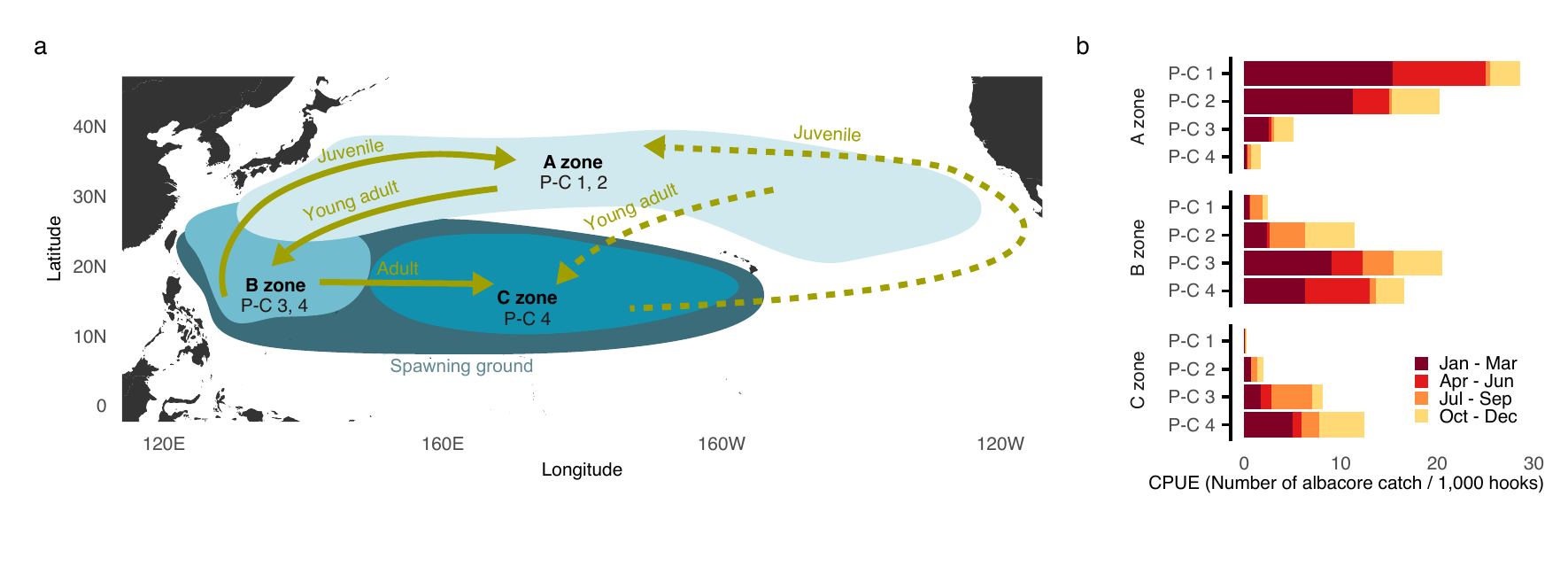}
\caption{\textbf{Albacore tuna migrations as suggested by our analyses.} \textbf{a,} Solid arrows indicate migration routes in the western north Pacific. Juvenile fish leave the spawning grounds to reach the A zone, where they seasonally congregate off the coast of Japan and disperse eastward. This lasts until the fish turn into late juveniles or early adults, when they relocate to the B zone. Later into their adulthood, the fish tend to move to the C zone that is placed more centrally in the north Pacific. Dashed arrows indicate possible migration routes in the eastern north Pacific, but the Japanese longline data is sparser here. Further investigations are needed for reliable results. \textbf{b,} Catch per unit effort (CPUE) spatially averaged across each of the zones highlighted in panel \textbf{a} reveals pseudo-cohort abundance that is compatible with the described migration routes. Importantly, the displayed routes should be seen as general tendencies for the albacore-tuna stock rather than as a linear progression in the life of any single fish.}
\label{f:mig}
\end{figure}

Taken together, the results show that albacore tuna prefer different habitats depending on their size and age. The seasonal appearance and disappearance of hotspots further points to likely migratory routes for albacore tuna, primarily in the western north Pacific ocean (Fig.~\ref{f:mig}a). Before discussing albacore-tuna migrations, however, we look at evidence that the identified hotspots, and thus major destinations for migrating fish, are indeed occupied by pseudo-cohorts as stated heretofore (Fig.~\ref{f:mig}b). To this end, we calculated the cohort-wise CPUE, spatially averaged over each of the three zones highlighted in Figure~\ref{f:mig}a. The A zone is dominated by the two smaller size pseudo-cohorts, with the January-March period being most abundant, followed by the April-June period (see the top panel in Fig.~\ref{f:mig}b). The two smaller size pseudo-cohorts almost disappear from this zone during the July-September period, only to reappear again in the October-December period. The B zone is primarily occupied by the two larger size pseudo-cohorts, with the January-March period again being most abundant, although other seasons substantially contribute to a complex overall pattern (see the middle panel in Fig.~\ref{f:mig}b). The second smallest pseudo-cohort is also abundant in this zone, especially in the period from July to December, thus substantiating the observation that smaller and larger size fish intermix here. Finally, the C zone is primarily occupied by the largest pseudo-cohort that is abundant throughout much of the year, followed by the second largest pseudo-cohort that is most abundant in the July-September period (see the bottom panel in Fig.~\ref{f:mig}b). The co-presence of the two larger size pseudo-cohorts substantiates the observation of fish intermixing within the zone's confines. 

\section*{Discussion}

Our study presents two main ecological results. First, we provide evidence that depending on size and age, albacore tuna prefer different habitats. Juveniles and early adults lean towards subtropical to temperate waters, whereas adults prefer tropical to subtropical waters thought to be the spawning grounds for the species. Second, we identify the likely albacore-tuna hotspots, and for each hotspot, provide the initial estimates of seasonal fish abundance. These results offer novel insights into albacore-tuna long-term migrations when the progression of habitat preferences is followed throughout ontogeny (Fig.~\ref{f:mig}a).

The study of albacore-tuna migrations has a long history, as exemplified by an early qualitative model of juvenile movements in the north Pacific ocean based on catch-and-effort data and conventional tagging by Otsu and Uchida~\cite{otsu1964model}. Using similar methodology, the eastward migration path towards the American west-coast fishery was examined by Laurs and Lynn~\cite{laurs1977seasonal}. These early results were quantitatively confirmed only much later~\cite{ichinokawa2008transoceanic}. Recent tagging programs provide detailed information on fish positions over time, but logistical and technical difficulties still limit the area of tagging and the period of recording; an example of this is archival tagging of north Pacific juvenile albacore tuna along the west coast of the United States, which returned 20 datasets between 63 and 697 days long~\cite{childers2011migration}. The focus of genetic studies, despite their potential to reveal fish migration pathways, has so far been more on the structure of global stocks~\cite{albaina2013single, montes2012worldwide, chow1995global}. The overall state of affairs emerging about the field is that of fragmented information which needs to be pieced together in extensive reviews to paint a holistic picture of long-term migrations spanning the entire albacore-tuna ontogeny~\cite{nikolic2017review}. Our results, by contrast, paint one such holistic picture for the western north Pacific ocean on their own (Fig.~\ref{f:mig}a). The situation in the eastern north Pacific remains somewhat opaque to us because the Japanese catch-and-effort data are naturally sparser in the east than the west.

To summarise key migration pathways (Fig.~\ref{f:mig}a), early juveniles leaving the western spawning grounds reach the A zone where they seasonally congregate off the coast of Japan and then disperse eastward in a pattern that appears consistent with the model of Otsu and Uchida~\cite{otsu1964model}. As these fish turn into late juveniles or early adults, they move to the B zone and start intermixing with other adult fish. Further growth prompts gradual shifting towards the C zone, placed more centrally in the north Pacific ocean. In relation to adults, our study offers details missing from the migration routes originating from reviewed literature~\cite{nikolic2017review}. Of note is that the described movements should not be seen as a linear progression in the life of a single individual. Rather, we are describing general tendencies that hold for the stock as a whole. 

Aside from deepening ecological understanding, our study has implications for various stakeholders, for example, fishers, conservationists, and managers. Considering which of these stakeholder groups is likely to receive the most utility from our findings, fishers are practitioners with hands-on experience with the albacore-tuna stock. They may, as is often the case with practitioners, already possess a good chunk of the knowledge laid out herein, albeit that knowledge need not be stored in a written form nor need it be easily articulated by any individual fisher. This leaves conservationists and managers for whom our findings may be helpful in devising less disruptive stock-management policies. The first step in this context could be scientific advice on which part of the albacore-tuna stock should be prioritised for conservation and management. In the case of a closely related species, Pacific bluefin tuna, population growth rate is most sensitive to juvenile survival~\cite{ijima2019effects}. This is because for prolific batch spawners like tunas, allowing early adults to go through a few reproductive cycles is sufficient to replenish the stock, despite the advantages that hyperallometric scaling of reproductive capacity may confer on more experienced adults~\cite{barneche2018fish}. If, accordingly, juveniles are prioritised for conservation and management, indiscriminate limits on the albacore-tuna fishery may be replaced with a policy such that the effort of vessels targeting smaller size pseudo-cohorts in the A zone is restricted.

From a more technical perspective on managing fish stocks, our study suggests a potential alternative to age-structured stock-assessment models that are commonly in use. Basing population dynamics on age structure is convenient, but ignores the fact that age is ecologically relevant only for very old individuals~\cite{vanleeuwen2010dynamic}. Animals, however, rarely live to old age in nature; they instead succumb to predation and diseases as the main causes of natural mortality. A good indicator of fish mortality by natural causes is body size~\cite{pauly1980interrelationships}. Fishing gear too is body-size, not age, selective. All this in turn means that age-structured models require some sort of a body size-age relationship, which is often given in the form of a von Bertalanffy growth curve (originally invented by Putter~\cite{kearney2021status}) and whose parameters are estimated from data using statistical methods. The parameters are thus static even in a changing environment, although environmental change is known to affect fish growth~\cite{jusup2014simple, cheung2013shrinking, sheridan2011shrinking}. These problems could be circumvented by switching to stage-structured models, but that raises the question of delineating ecophysiologically distinct cohorts. Our study suggests that we could perhaps let the data speak for itself and use a data-driven approach to learn pseudo-cohorts. Doing so would also resolve the problem of defining fisheries, laid out in the Introduction section, because there is one-to-one correspondence between fisheries and pseudo-cohorts when the latter are learnt from catch-and-effort data. The approximate age of pseudo-cohorts would still have to be estimated, but that would involve relatively modest sampling of age-revealing structures such as otoliths~\cite{wells2013age, williams2013comparison}. Overall, a range of technical difficulties would undoubtedly follow any attempt at implementing stock assessments using a modelling approach that substantially differs from the current common practices. We posit, however, that rewards are likely to outweigh the difficulties, perhaps even by a large margin.

\section*{Methods}

\paragraph*{Dataset.} We analysed the Japanese longline logbook data for the period from 1994 to 2020, containing records on 889,391 operations in the north Pacific executed over the course of 86,907 voyages. Longline logbooks collect detailed information about fishing operations including dates, locations, vessel names and categories, voyage numbers, fishing-gear configuration, the number of fish caught, catch mass, and effort in number of hooks~\cite{ward2007overview}. Longline vessels typically cast nylon ropes with thousands of hooks into the ocean and catch top pelagic predators such as tunas, billfishes, and sharks. Albacore tuna are one of the target species for longline fishery. Anywhere between zero and more than a hundred individuals can be caught over the course of a single operation. Fishers take fish migrations into consideration and adjust their gear configuration to different targets depending on the season and fishing location~\cite{miyabe1987review}. Accordingly, the size of caught fish varies seasonally and geographically. 

Individual fish masses were not available in the dataset. Instead, we used the average body mass of fish caught during a single fishing operation, which we obtained by dividing the single-operation catch mass by the total number of fish caught. Single-operation catch mass is itself also an estimate recorded by the fishing crew. In about 17\,\% of instances only the whole-trip average body mass was available. The average body mass is less precise than direct port sampling, but port-sampling records are sparse and cover a much narrower geographic range.

The Japanese longline fishery comprises vessels of various capacities. Based on their tonnage-dependent licences, vessels are legally classified into three categories: coastal, offshore, and distant-water. Categorisation relates to the way vessels operate, for example, their range and the number of days at sea. Distant-water vessels are thus equipped with deep freezers, allowing them to make long, far-sea voyages. Coastal longliners are, by contrast, confined to a stipulated operational area. Importantly, we made use of logbook data from all vessel categories in order to incorporate information from as wide an ocean area as possible.

\paragraph*{Finite mixture modelling.} We assumed that body masses used in our analyses originate from a multimodal, mixture distribution. The assumption rests on the grounds that the albacore-tuna stock consists of cohorts that are sufficiently distinct from one another both from an ecological perspective and the perspective of albacore-tuna fishery. We constructed a finite mixture model~\cite{mclachlan2019finite} to separate the starting mixture distribution into constituent (also called latent or marginal), monomodal distributions corresponding to above said cohorts.

The probability density function of a mixture distribution with $K$ constituents can generally be written
\begin{linenomath}
\begin{equation}
p(x) = \sum^{K}_{k=1} \pi_{k} f(x | \bm{\theta}_{k}),
\end{equation}
\end{linenomath}
where $f$ is a monomodal probability density function, $\pi_{k}$ are mixing coefficients, and $\bm{\theta}_{k}$ is the parameter vector for the function $f$. In our dataset, body masses are continuous, positive values that occasionally get large enough for the constituent, monomodal distributions to be moderately right-skewed. We therefore chose to work with the gamma distribution whose parameters are $\alpha$ and $\beta$ such that $f(x | \bm{\theta}_{k}) = \Gamma(x | \alpha_{k},\beta_{k})$. 

A few additional details needed to be addressed before specifying the likelihood function for the problem at hand. The first detail was that body masses depend on location and time, meaning that all fish caught during adjacent fishing operations are repeated measures rather than independent observations. Accordingly, year, month, and location are grouping factors in the model, where location is specified on a 1$^\circ$$\times$1$^\circ$ latitude-longitude grid. The second detail was that albacore-tuna life cycle is seasonal~\cite{nikolic2017review}, implying that cohort characteristics change throughout the year. We thus decided to separate logbook data into four seasons or quarters: January-March, April-Jun, July-September, and October-December. If the number of observation groups is $G$ and the number of observations in group $g$ is $N_g$, then the log-likelihood function is
\begin{linenomath}
\begin{equation}
\log{\mathcal{L}} = \sum^{G}_{g=1} \sum^{N_g}_{n=1} \log\left(\sum^{K}_{k=1} \pi_{k} \Gamma(x_{g,n} | \alpha_{k},\beta_{k})\right).
\end{equation}
\end{linenomath}

We used the R software package \texttt{FlexMix} for the analysis. The package employs expectation-minimisation (EM) algorithm to maximise the likelihood function $\mathcal{L}$~\cite{leisch2004flexmix, grun2008flexmix}. We varied the number of potential constituent, monomodal distributions $K$ from 1 to 6, while relaying on the Bayesian information criterion (BIC) to decide on the appropriate value of $K$. We also checked lower-resolution grouping factors such as year and location, or location only, but the BIC indicated a poorer performance. The monomodal constituents of the starting mixture distributions are referred to as pseudo-cohorts, where the prefix `pseudo' signifies a data-driven definition in place of an ecophysiological one. For each pseudo-cohort, we calculated spatially explicit catch per unit effort (CPUE) as an indicator of pseudo-cohort abundance in the north Pacific.

%\nolinenumbers

%\bibliography{sample}

\section*{Acknowledgements}

We thank Steven Teo, Sarah Hawkshaw, and Jun Matsubayashi for their comments on the study. This work was supported by the Research and assessment program for internationally managed fisheries resources, the Fisheries Agency of Japan. M.J. was partly supported by the Japan Society for the Promotion of Science (JSPS) KAKENHI grant no. JP21H03625. 

\section*{Author contributions statement}

H.I.\ wrote the code. H.I.\ and M.J.\ designed the study and analysed the results. All authors discussed the results and reviewed the manuscript.

\section*{Additional information}

Authors declare no conflict of interest. 

\end{document}